\documentclass[12pt,preprint]{aastex}
\usepackage{epsfig}

\shorttitle{Galaxy Structure in the UV}
\shortauthors{Mager, Conselice, Seibert et al.}

\begin{document}

\title{Galaxy Structure in the Ultraviolet: The Dependence of Morphological
Parameters on Rest-Frame Wavelength}

\author{Violet A. Mager\altaffilmark{1},
Christopher J. Conselice\altaffilmark{2},
Mark Seibert\altaffilmark{3},
Courtney Gusbar\altaffilmark{4},
Anthony P. Katona\altaffilmark{5},
Joseph M. Villari\altaffilmark{5},
Barry F. Madore\altaffilmark{3},
\& Rogier A. Windhorst\altaffilmark{6}}


\altaffiltext{1}{Pennsylvania State University Wilkes-Barre, Old Route 115, Lehman, PA 18627}
\altaffiltext{2}{University of Nottingham, School of Physics and Astronomy, University Park, Nottingham NG7 2RD, UK}
\altaffiltext{3}{Carnegie Observatories, 813 Santa Barbara St., Pasadena,
CA 91101}
\altaffiltext{4}{Ohio University, Department of Physics and Astronomy,
Athens, Ohio 45701}
\altaffiltext{5}{Susquehanna University, Department of Physics, 514
University Avenue, Selinsgrove, PA 17870}
\altaffiltext{6}{Arizona State University, School of Earth \& Space Exploration, Tempe, AZ 85287-1404}

\clearpage
\begin{abstract}
Evolutionary studies that compare galaxy structure as a function of redshift
are complicated by the fact that any particular galaxy's appearance depends 
in part on the rest-frame wavelength of the observation. This leads to the 
necessity for a "morphological k-correction" between different pass-bands, 
especially when comparing the rest-frame optical or infrared (IR) to the 
ultraviolet (UV). This is of particular concern for high redshift studies
that are conducted in the rest-frame UV. We investigate the effects of this 
"band-pass shifting" out of the UV by quantifying nearby galaxy structure 
via "CAS parameters" (concentration, asymmetry, and clumpiness). For this 
study we combine pan-chromatic data from the UV 
through the near-IR with GALEX (Galaxy Evolution Explorer) data of 2073 
nearby galaxies in the "NUV" ($\sim 230$ nm) and 1127 in the "FUV" 
($\sim 150$ nm), providing the largest study of this kind in the mid- to 
far-UV. 
We find a relationship between the CAS parameters and observed
rest-frame wavelength that make galaxies appear more late-type 
at shorter wavelengths, particularly in the UV. The effect is strongest
for E/S0 galaxies in the far-UV, which have concentrations
and asymmetries that more resemble those of spiral and peculiar/merging
galaxies in the optical. This may be explained by 
extended disks containing recent star formation. Here we also release
the CAS values of the galaxies imaged in GALEX NUV and FUV for use 
in comparisons with deep HST imaging and JWST in the future.
\end{abstract}


\section{Introduction}

The majority of high redshift (z $> 2$) galaxies appear similar to a relatively 
rare subset of low redshift irregular and peculiar galaxies, whose morphologies 
appear to be pathological due to mergers or interactions 
(e.g., Driver et al. 1995,
Conselice et al. 2005, Kriek et al. 2009, Delgado-Serrano et al. 2010,
Mortlock et al. 2013). This observed increase in the percentage of 
apparently merging/interacting galaxies with redshift supports models of 
hierarchical galaxy formation (e.g. Nagashima et al. 2002). 

Comparison studies of galaxies at different redshifts are complicated, 
however, by the fact that galaxies can appear substantially different at 
shorter wavelengths than at longer ones (e.g., Bohlin et al. 1991, 
Kuchinski et al.  2000, Windhorst et al. 2002). This is especially true 
for galaxies that appear as
earlier types at the longer wavelengths. This leads to a "morphological 
k-correction" for 
a given galaxy between different rest-frame wavelengths. This is 
particularly important in studies of high redshift galaxies, as band-pass 
shifting will cause light originally emitted in the UV to be shifted as far 
as the IR by the time it reaches Earth. This raises questions about how much 
of the irregular/peculiar morphologies seen in high redshift studies are 
simply due to band-pass shifting, and not due to real differences in galaxy 
type.
Kuchinski et al. (2001) showed that this effect alone is enough to 
misclassify galaxies as peculiar when simulated at higher redshifts. A 
quantitative measurement of the morphological k-correction is therefore 
an essential first step to the study of galaxy evolution. 

This correction is expected to be particularly significant when comparing
the rest-frame UV to the optical or IR, as galaxy stellar energy 
distributions (SED's) can change drastically short-ward of the Balmer Break 
($\lambda_c \lesssim 360$~nm), and UV-bright star-forming regions dominate 
galaxy morphology that appears to be smoother at redder wavelengths. 
To determine the 
extent of this effect, it is critical to conduct a large representative 
benchmark study at $z \simeq 0$. This is complicated by the difficulty 
of observing in the UV through the atmosphere: it is necessary
to use space-based telescopes with UV sensitive detectors, such as HST
(Hubble Space Telescope) and GALEX (Galaxy Evolution Explorer). Throughout
its operation, GALEX has amassed the largest image database in the far-UV
to date (Bianchi \& GALEX team 1999, Martin \& GALEX team 2005, 2009). 
This makes GALEX ideal for these types of studies, where
large number statistics are needed to determine overall generalized trends in
galaxy structure.
 
One of the most efficient ways to quantify the morphological k-correction is
through measurements of the concentration, asymmetry, and clumpiness (CAS) 
parameters of galaxies as a function of their rest-frame wavelength. The 
concentration of light as it is radially distributed within a galaxy (C)
correlates with the stellar and kinematic masses of galaxies. The degree to
which a galaxy deviates from perfect symmetry (A) can indicate
on-going galaxy mergers or interactions. The proportion of high frequency
structure to the smooth light distribution of a galaxy (S) correlates with 
current star formation (Conselice 2003). Conselice (2006) argues that 
these parameters are the more fundamental properties that describe a 
galaxy's physical state, and thus are a good objective way of classifying 
galaxies. Nearby galaxy studies have shown that a combination of these 
parameters provides a relatively robust automated galaxy classification 
system, which can be used to determine the nature of the population 
distributions for large samples of galaxies (e.g. Bershady et al. 2000,
Conselice 2003, Conselice 2006). In Taylor-Mager et al. (2007), we found that 
earlier-type galaxies are on average increasingly more concentrated, 
more symmetric, and less clumpy than later-type galaxies. Merging galaxies 
occupy separate locations within the CAS parameter space, which leads to a 
clear and efficient method of automatically identifying merging galaxies. 

There have been a few studies that quantify the morphological k-correction
of nearby (z $\sim 0$) galaxies (e.g., Bohlin et al. 1991, 
Windhorst et al. 2002, Vika et al. 2013), one of the largest of which is our 
earlier project (Taylor-Mager et al. 2007) that utilized 
multi-wavelength (0.15-0.85$\mu$m) CAS measurements of 199 galaxies. Only
a subset of this sample had UV data at the time, with GALEX observations for 
just 14 of them. Due to signal-to-noise (S/N) and resolution constraints, 
only 7 galaxies were used to calculate the final k-correction in the 
NUV ($\lambda_c=2275$\AA) and 5 in the FUV ($\lambda_c=1550$\AA). 
Complimentary HST WFPC2 data allowed for measurements at longer UV 
wavelengths, with 8 galaxies in the F255W ($\lambda_c=2550$\AA) filter and 86 
in F300W ($\lambda_c=2930$\AA). 
Galaxy number statistics were small in the UV
for all galaxy types, but particularly for early-type (E--S0) galaxies, which
were not the focus of that study, and which are especially faint in the
UV due to their intrinsically red colors. As such, we were unable to 
provide a reliable quantified measure of the morphological k-correction in 
the UV for E--S0 galaxies in that paper. We noted an intriguing trend
of E--S0 galaxies apparently becoming dramatically less concentrated 
in the UV, but included the caveat that this needed to be verified with a 
larger sample of early-type galaxies.

This has motivated our present study, which uses images in the GALEX 
archive to
drastically improve our number statistics (by up to almost a factor of 300)
in the far- to mid-UV, providing the largest analysis of the 
wavelength-dependence of nearby galaxy CAS structure in this wavelength 
range to date.

\section{Data Analysis}
Conselice (2003) and Conselice et al. (2000) showed that poor spatial 
resolution and low signal-to-noise ratio (S/N) affect the reliability of CAS 
measurements. As such, we applied the selection criteria of  
S/N $\geq 75$ and linear resolution $\leq 1.25$ kpc to our sample.
These limits are more relaxed than the upper limits determined in 
Conselice et al. (2000) (S/N $\geq 100$ and resolution $\leq 1$ kpc), 
but were chosen 
to be consistent with the combined ground-based and HST sample
(Taylor-Mager et al. 2007), for which we determined that these 
relaxed limits did not have a significant impact on the CAS measurements,
and allowed for improved number statistics.  Our resolution-limited
sample contains 2967 galaxies in the GALEX GR6 data release\footnote{
galex.stsci.edu/GR6} that have images
in at least the NUV filter. Applying the signal-to-noise cut (see below
for details on how it was calculated) 
reduced the sample to 2073 galaxies in NUV and 1127 galaxies in the FUV.
A histogram of S/N values is shown in Figure 1. Figure 2 displays
example images of galaxies with three different S/N ratios in NUV. The
top left galaxy meets our relaxed requirement of S/N $\geq 75$, but falls 
short of the more strict requirement of S/N $\geq 100$. The top right
galaxy has S/N just over 100, and the bottom galaxy has a particularly
high S/N of over 1000.

We processed the images using the pipeline designed by the GALEX team for 
creating extended source catalogs (Seibert et al. 2012), which are published
as the GALEX All-Sky Survey Source Catalog (GASC) and Medium Imaging Survey
Source Catalog (GMSC). Parameters such as the position of the galaxy center,
size, ellipticity, and inclination were found using the pipeline's automated 
software, and
then verified by-eye, with corrections made as necessary if the target
galaxy was not properly parameterized by the automated algorithm. 
Non-target sources bright and close enough to the 
target galaxy to significantly affect photometry were manually identified and 
marked for inclusion in a mask file. In this initial step, we removed only 
these sources from our images by interpolating over them.
From the panchromatic object counts of Windhorst et al. (2011,
their Fig. 11 and 12), we expect an integrated 10 $galaxies/deg^2$ 
to AB $< 15$ mag in V-band. From Windhorst et al. (2008),
the median half-light radius $r_{hl} < 0.5$ arcmin at V $< 15$ mag, so the area
in the sky covered by galaxies with V $< 15$ mag is $< 7.8$ $arcmin^2$. The
star counts of Windhorst et al. (2011, their Fig. 11) have an
integrated 200-500 $stars/deg^2$ (valid for a range of $b_{II}$) to V $< 15$
mag, or $< 0.056-0.14/arcmin^2$. The areal ratio of stars/galaxies
in the sky to AB $< 15$ mag is therefore $< 0.7-1.8\%$, i.e. generally less
than $2\%$ of nearby galaxies will have been affected by a Galactic star.
Also, only a few percent of nearby galaxies have close companions
(e.g., Mantha et al. 2018). Hence, no more than a few percent of all
target galaxies will have been affected by stars or neighboring
galaxies, which were masked out in the standard GALEX processing
stage.

All other non-target sources were excluded while measuring CAS parameters
as described in the next paragraph. The
signal-to-noise (S/N) of each galaxy was calculated using the following
formula, the image exposure time (exptime), the total object counts per 
second (int) above the local sky-value, and the background counts per 
second (bg)., as measured by the GALEX pipeline process
(Seibert et al. 2012).

\begin{equation}
S/N = {int \times exptime \over \sqrt{exptime \times (int+bg)}}
\end{equation}

In our previous study (Taylor-Mager et al. 2007), we removed all 
non-target sources from our images before feeding them into the automated 
IRAF\footnote{IRAF is distributed by the National 
Optical Astronomy Observatories, which are operated by the Association 
of Universities for Research in Astronomy, Inc., under cooperative 
agreement with the National Science Foundation.} 
task CAS, developed by C. Conselice (Conselice et al. 2000, 
Conselice 2003). In this earlier version of the CAS software, a region of empty 
sky was chosen for background calculations.
For this study, we use an improved version of the CAS software, which uses
a more accurate method of measuring the sky. Source segmentation
maps are produced by SExtractor (Bertin \& Arnouts 1996) and used by the 
CAS code to remove contaminating objects, replacing them with a local sky 
background including its proper noise characteristics. We used this CAS code 
to measure the CAS parameters as defined by Conselice et al. (2000) and
Conselice (2003), as described below.

The concentration index, C, is computed by determining the total
sky-subtracted light contained within $1.5 \times$ the galaxy's Petrosian 
radius (r$_{Pet}$) (Petrosian 1976), 
and finding the logarithmic ratio of
the radius within which 80\% of this light is contained (r$_{80}$) to the 
radius within which 20\% of this light is contained (r$_{20}$)
(Conselice et al. 2000).
The concentration index is thus given by the formula:
\begin{equation}
C = 5 \times log(r_{80}/r_{20})
\end{equation}
Galaxies that are highly concentrated in their centers will have high 
values of C, and galaxies that have less-centrally concentrated light 
distributions will have lower values of C.

The asymmetry index, A, is computed by rotating the galaxy by 
180\degr~from its center, and subtracting the
light within $1.5 \times$ r$_{Pet}$ in the rotated image (I$_{180}$) 
from that in the original image (I$_o$). The ratio of the residual 
subtracted flux to the original galaxy flux yields A:
\begin{equation}
A = min ({\Sigma~| I_o - I_{180} | \over {\Sigma~| I_o |}}) -
min ({\Sigma~| B_o - B_{180} | \over {\Sigma~| I_o |}})
\end{equation}
Noise corrections are applied by subtracting the asymmetry of the sky 
(B), and iterative centering corrections are applied 
to minimize A, since gross centering errors would artificially increase
the derived A-values. The center used to minimize A was not used for the
concentration index, which instead utilized the centroid of light. 
Conselice (2003) found a difference of at most a few pixels between these
centers, or roughly $2\%$ the size of the galaxy, and the center choice had
no effect on the determination of C, even to within $5-10\%$ of the size
of the galaxy. Asymmetries range from A = 0 to 2, with A = 0 
corresponding to a truly symmetric galaxy, and A = 2 corresponding to
a very asymmetric galaxy.

The clumpiness index, S, is defined as the ratio of the amount of light
in high spatial frequency structures within $1.5 \times$ r$_{Pet}$ to the 
total amount of light in the galaxy within that radius (Conselice 2003).
This is done by subtracting a boxcar-smoothed image from the original image
to produce an image that contains {\it only} its high-frequency structure.
The central pixels within 1/20th of the total radius ($1.5 \times$ r$_{Pet}$) 
are set to zero to exclude them from the parameter measurements.
S is given by the following equation, where I$_{x,y}$ is the intensity 
of light in a given pixel, I$^s_{x,y}$ is the intensity of 
that pixel in the image smoothed by $0.3 \times $r$_{Pet}$, 
and B$^s_{x,y}$ is an intensity 
value of a pixel from a smoothed background region: 
\begin{equation}
S = 10 \times ({\Sigma^{N,N}_{x,y=1,1}~(I_{x,y} - I^s_{x,y}) \over 
\Sigma^{N,N}_{x,y=1,1}~I_{x,y}} - 
{\Sigma^{N,N}_{x,y=1,1}~B^s_{x,y} \over \Sigma^{N,N}_{x,y=1,1}~I_{x,y}})
\end{equation}
A clumpiness of S = 0 corresponds to a galaxy that has no high frequency
structure, and is therefore completely smooth.
Galaxies with more high-frequency structure are more patchy in appearance,
and so will have a higher value of S. 


\section{Results}
\subsection{GALEX CAS Analysis}

Table 1 lists the galaxies in our sample and their CAS parameters as measured
in the NUV and FUV, for which S/N $\geq 75$. Table 2 lists CAS parameters for
galaxies that had S/N $< 75$ in the NUV and/or FUV. The first 10 galaxies 
are listed here, with the full
tables available in the electronic edition. Types were obtained
from the homogenized galaxy morphology classification listed in NED (NASA/IPAC
Extragalactic Database), and each galaxy was placed into a numerical type
category defined as follows: 1 = E-S0 (elliptical and lenticular), 
2 = Sa-Sc ("early-type" spirals), 3 = Sd-Im ("late-type" spirals and
irregulars), and 4 = peculiar or merging. All galaxies that had a peculiar
(pec) designation in their classification were placed in the 4th type bin,
regardless of any other designation they may have also had.
Galaxies that could not be placed into any of these categories were 
given no type (such as those without any classification listed in NED, or 
non-specific classifications such as S?). 

We plot each galaxy in "CAS space" (concentration and clumpiness versus
asymmetry) in the left panels of Figure 3 for the NUV, and Figure 4 for the 
FUV, with symbols coded by galaxy type. Median values with their errors 
(solid error bars) and 25-75\% quartile ranges (dotted error bars) for each 
type bin are plotted in the right panels. Median
values were calculated from 134 E/S0, 852 Sa-Sc, 600 Sd-Im, and 312 Pec 
galaxies in
the NUV, and 56 E/S0, 485 Sa-Sc, 364 Sd-Im, and 157 Pec galaxies 
in the FUV, when
using the criteria that S/N $\geq 75$. 

We tested how the median values would
be affected by a stricter limit of S/N $\geq 100$, and found no difference
within the errors for all but two of the data-points
(NUV C for Sa-Sc, and FUV A for Sd-Im), which differed (S/N cut at 100 minus
S/N cut at 75) by $-0.03 \pm 0.02$, a $1.5\sigma$ difference. The number
of E/S0 galaxies was lowered from 134 to 106 in the NUV, and 56 to 37 in 
the FUV.
Since there was little-to-no effect on our results by using the stricter
S/N constraint, to remain consistent with the HST/VATT data and preserve
higher number statistics for the under-represented E/S0 galaxies, we continue
with the S/N $\geq 75$ constraint.

As seen in Figures 3 and 4, while there is a significant amount of 
overlap between galaxy types, there is
an overall average trend of galaxies in both pass-bands tending to appear
{\it less concentrated}, {\it more clumpy}, and {\it more asymmetric} 
when progressing from "early-type" (E/S0) through "late-type" (Sd-Im) 
galaxies on the Hubble sequence.
This same effect is also apparent in the near-UV through near-IR
(Taylor-Mager et al. 2007), and coincides with the expectations, since
later-type galaxies are known to have lower bulge-to-disk ratios 
(and thus a higher percentage of light in the extended outer regions) 
and more recent star formation (which contributes to clumpy regions and an 
overall asymmetry).

The peculiar galaxy category contains many different types of galaxies,
from full-on major mergers to otherwise normal elliptical or spiral galaxies 
that contain some level of unusual structure. It is therefore not surprising
that their median values in Figures 3 and 4 fall near the average of all 
galaxy types, although they do tend to be slightly more concentrated than 
the average normal spiral. In Taylor-Mager et al. (2007), we found a 
progression through merger stages, where galaxies initially become 
significantly less concentrated and more asymmetric and clumpy, then settle 
back closer to the locus of normal galaxies as post-mergers. The 
concentration indices of these post-mergers, however, are notably higher 
than the average normal spiral galaxy, as gas was funneled into the center 
by the merger process. This suggests that our peculiar category is 
dominated by post-mergers, with the presence of visible tidal or
"disturbed" features leading to the "peculiar" classification.

Comparing the median values in Figures 3 and 4 reveal that E/S0 
galaxies are on average significantly less concentrated and more asymmetric
in the FUV than in the NUV, and later galaxy types are significantly more 
asymmetric and clumpy in the FUV than in the NUV.

\subsection{CAS Wavelength Dependence}

Figure 5 shows the relationship of the median CAS parameters for each
type bin with observed rest-frame wavelength. Here we supplement the FUV and
NUV results from this study with data used in our previous study 
(Taylor-Mager et al. 2007), which includes HST WFPC2 mid- to near-UV 
(F255W, $\lambda_c = 255$ nm and F300W, $\lambda_c = 293$ nm) and near 
IR data (F814W,
$\lambda_c = 823$ nm), as well as ground-based UBVR
($\lambda_c = 360$, 436, 540, and 634 nm, respectively) data
from the Vatican Advanced Technology Telescope (VATT). Solid error bars
indicate the error on the median, while dashed error bars give the
25-75\% quartile ranges, which indicate the spread of the data. The
number statistics in the FUV and NUV are vastly improved in this study,
which has resulted in much smaller error bars for each of the spiral, 
irregular, and peculiar galaxy type categories. 

For Figure 5 we re-processed the Taylor-Mager et al. (2007) data through
the newest version of the CAS code, as it treats background removal
differently (utilizing the entire image with objects masked out via
SExtractor catalogs, vs. a small user-defined box of empty
sky). This had a minor impact on a small number of data-points as compared to
our 2007 results from the previous version of the code, but the overall trends 
with wavelength remain the same. This demonstrates that
the difference in background subtraction methods does not have a
significant impact on our conclusions.
We rejected any median data-points derived from fewer
than 5 HST galaxies, as these low number statistics would produce
unreliable results.

Most panels in Figure 5 show a consistent general trend in CAS
values as a function of rest-frame wavelength. A major exception is the 
clumpiness (S) 
parameter for all galaxy types except E/SO, where S increases signifcantly
with decreasing wavelength, then drops dramatically between the HST/VATT
UV and the shorter GALEX NUV wavelength (the second data-point from the left in
each panel). The clumpiness parameter is particularly sensitive to
the image resolution, as the high-frequency structure measured by the S 
value is 
smoothed out in low resolution images. The GALEX images (left 2 data-points
in each panel) have a stellar FWHM of 5 arcsec, while the VATT and the
convolved HST images from Taylor-Mager et al. (2007) have an average stellar
FHWM of 1.75 arcsec.\footnote{For Taylor-Mager et al. (2007), all images
were convolved with a differential PSF, such that the total PSF in all
images had FWHM $\sim 1.75$. In the current study we do not convolve all
data to the course GALEX FWHM, in order to better see the optical trends.}

To test the effect of image resolution, we convolved
the HST F300W images to the GALEX resolution and measured the resulting
median CAS values. We found no change within the uncertainties for
any of the E/S0 CAS values, and no change in C and A for the peculiar 
galaxies, and C for Sa-Sc. There was a minor ($1.6 \sigma$) decrease
in A for Sa-Sc and Sd-Im at the poorer resolution, and a $2.1 \sigma$ increase
in C for Sd-Im. As expected, the S values were generally impacted the most, 
with a decrease of $3.4\sigma$, $1.2\sigma$, and $2.3\sigma$ for Sa-Sc, Sd-Im, 
and pec/mer, 
respectively. We therefore conclude that there is an artificial
systematic baseline difference due to differences in resolution for some
of the panels in Figure 5. This explains the discrepancy in wavelength
dependence between the GALEX NUV and HST F300W or VATT U-band data, as
removing the effect would result in GALEX measurements that are significantly 
higher in S for all but the E/S0 galaxies (Sa-Sc: $0.48 \pm 0.14$,
Sd-Im: $0.06 \pm 0.05$, and pec/mer: $0.18 \pm 0.08$), slightly higher in A for
Sa-Sc ($0.21 \pm 0.13$) and Sd-Im ($0.14 \pm 0.09$), and slightly lower 
in C for Sd-Im ($-0.17 \pm 0.08$). As these corrections were measured
in F300W, we do not apply them to the GALEX data in Figure 5, but consider
their effects in our discussion of the results below.

Early to mid-type spirals (Sa-Sc) are generally less concentrated 
and more asymmetric and clumpy at increasingly shorter observed wavelengths.  
Although the two shortest wavelength data-points (GALEX NUV/FUV) have much
lower clumpiness than expected from the trend displayed at longer 
wavelength, they exhibit the same behavior of increasing clumpiness
with decreasing wavelength. The systematically lower baseline difference
can be explained by the relatively poor resolution of the GALEX
data. The late-type spiral, irregular, and peculiar galaxies show the
same trends in A and S, but a weaker relationship in C, where the
concentration index changes little overall with observed wavelength.

Our previous study (Taylor-Mager, et al. 2007) contained too few 
E/S0 galaxies in FUV and NUV to include significant data-points on this plot. 
In that study, the optical/near-IR data showed little-to-no 
indication of a
dependence of CAS parameters on wavelength for E/S0 galaxies, but the HST
near-mid-UV indicated a sudden, drastic decrease in C and a mild 
increase in A, when transitioning toward shorter wavelengths in the UV. 
However, the
uncertainties on the HST data were relatively large, and they were based
on few galaxies (2-4 per median data-point). As such, we included the
caveat that these rather surprising conclusions were only tentative.

The superior number statistics and smaller uncertainties of this GALEX study
confirm an interesting transition in the morphologies
of E-S0 galaxies when viewed in the mid- to far-UV.
As seen in Figure 5, E/S0 galaxies are significantly less concentrated,
more asymmetric, and slightly more clumpy than at redder 
colors, where the CAS parameters do not vary much with wavelength. Tests
with the F300W data show that the clumpiness parameter for the
inherently smooth E/S0 galaxies is not significantly affected by resolution, so 
we do not expect much of a correction for the low resolution far-UV
GALEX data.
This dramatic change in A, and the more mild change in S was not 
apparent from the Taylor-Mager et al. (2007) HST UV data.

Table 3 lists the differences in the median CAS parameters between the NUV and
the FUV for each type bin ($\delta$) and the errors on these differences
($\sigma$). This provides a quantitative correction between these structural
parameters as measured at these two rest-frame wavelengths. These differential
values show that there is a very significant drop in concentration
for E/S0 galaxies when comparing the FUV to the NUV 
($\delta_C (E/S0) = 0.45 \pm 0.11$). This is more extreme than the drop in C
for early-mid type spirals ($\delta_C (Sa-Sc) = 0.10 \pm 0.03$). Within the 
uncertainties, there is no change in C for late-type spirals, irregulars, and
peculiars. All galaxy types are significantly more asymmetric in the FUV
than the NUV, with the largest difference being for early to mid-type spirals. 
Spiral, irregular, and peculiar galaxies are all significantly more clumpy
in the FUV than the NUV, but E/S0 galaxies are only more clumpy by slightly 
more than $1\sigma$. 

\section{Conclusions}

In conclusion, we have measured the concentration, asymmetry, and
clumpiness (CAS) parameters of 2073 nearby galaxies imaged in the far-UV by
GALEX, and release those values for use in future comparison studies, such
as those with deep HST and JWST images.

We find a significant morphological k-correction in C
for E/S0 and Sa-Sc galaxies, with decreasing concentration at shorter
wavelengths. There is a particularly dramatic difference in concentration
for E/S0 galaxies between the far- and mid-UV. Sd-Im and peculiar/merging
galaxies show a weaker trend in C, where concentration does not vary much
between filters. There is a strong morphological k-correction in
A for all galaxy types, with increasing asymmetry at shorter wavelengths.
However, there is little change in A for E/S0 galaxies within optical
and IR wavelengths.
There is also a strong k-correction in S of increasing clumpiness at
shorter wavelengths for spirals, irregulars, and 
mergers/peculiars. For these galaxy types, we find that care should be taken
when comparing images of vastly different spatial resolutions, as much lower
resolution leads to a much lower clumpiness value. This resolution effect is 
not as strong for the inherently smooth E/S0, for which we find little
dependence of S on wavelength, except for a mild increase in clumpiness
when transitioning from optical to UV wavelengths. The A values, on the other
hand, are not significantly affected by image resolution, which show that 
asymmetries
are large scale features that do not disappear at the lower (5 arsec 
stellar FWHM) 
GALEX resolution.

The relationship between the CAS parameters and the observed rest-frame
wavelength makes galaxies in general appear more late-type than they really
are at shorter wavelengths, especially in the mid- to far-UV, where
the morphologies with CAS become nearly degenerate for all but the early type 
galaxies (as apparent in Figures 3 and 4).
Surprisingly, this effect is so strong for E/S0 galaxies in the far-UV that 
their
concentrations and asymmetries more closely resemble those of spirals
and peculiar/merging galaxies at red optical wavelengths.
These high asymmetries and low concentrations of ellipticals and
lenticulars at shorter wavelengths
can be explained by extended ultraviolet disks and halos that have been found
in many spiral and irregular galaxies, where low-level star
formation is ongoing in the faint outskirts (e.g. Holwerda
et al. 2012, Hodges-Kluck et al. 2016). Most surprisingly, early-type 
(E/S0) galaxies historically considered to be "red and dead" 
have been found to be UV-bright (Schawinski et al. 2007), many with 
extended disks visible primarily in the UV 
(Rutkowski et al. 2012, Petty et al. 2013). This accounts for their diffuse and asymmetric structure
as measured here in the UV, resulting in morphologies that more resemble 
those of disk galaxies in the optical. This UV excess in 
early-type galaxies can potentially be explained by recent star formation,
or an evolved horizontal 
branch stellar population (O'Connell 1999; "UV upturn").  
The stellar population in the inner regions of some E/S0 galaxies have been
found to be older than the outer regions (Petty et al. 2013), with more 
recent star formation in extended H I-rich regions (Yildiz et al. 2017).
In at least some cases this may be the result of recent interactions with
companion galaxies or gas-rich merger events (Koshy \& Zingade 2015).

\acknowledgments

We thank the peer reviewer for his or her careful review of our paper.
This research has made use of the NASA/IPAC Extragalactic Database
(NED) which is operated by the Jet Propulsion Laboratory, California
Institute of Technology, under contract with the National
Aeronautics and Space Administration. This research has also made use of
NASA's Astrophysics Data System. The research was funded by NASA
grant NNX09AF83G. RAW is supported by NASA JWST  
Interdisciplinary Scientist grants NAG5-12460, NNX14AN10G,
and 80NSSC18K0200 from GSFC.

{\it Facilities:} \facility{GALEX} \facility{HST}

\noindent
\begin{deluxetable}{llllllllll}
\tabletypesize{\scriptsize}
\tablecolumns{10}
\tablewidth{0pt}
\tablecaption{High S/N Measured CAS Parameters} 
\tablehead{
\colhead{Galaxy} & 
\colhead{R.A. (deg)} & 
\colhead{DEC (deg)} &
\colhead{Type} & 
\colhead{C$_{NUV}$} & 
\colhead{A$_{NUV}$} & 
\colhead{S$_{NUV}$} & 
\colhead{C$_{FUV}$} &
\colhead{A$_{FUV}$} &
\colhead{S$_{FUV}$}
\cr
\colhead{\nodata} &
\colhead{S/N$_{NUV}$} &
\colhead{S/N$_{FUV}$} &
\colhead{\nodata} &
\colhead{$\sigma_{C_{NUV}}$} &
\colhead{$\sigma_{A_{NUV}}$} &
\colhead{$\sigma_{S_{NUV}}$} &
\colhead{$\sigma_{C_{FUV}}$} &
\colhead{$\sigma_{A_{FUV}}$} &
\colhead{$\sigma_{S_{FUV}}$}
}
\startdata

2MASXJ11102621+5824080 & 167.60939026 &  58.40214920 & 2 &  3.066 & 0.189 & 0.200  & 9.999 & 9.999 & 9.999\\
\nodata & 88.3 & 30.4 & \nodata & 0.163 & 0.047 & 0.018 & 9.999 & 9.999 & 9.999\\
ESO013-016  &  23.19899900 & -79.47364800 & 3 &  2.313 & 0.554 & 0.260 & 9.999 & 9.999 & 9.999\\
\nodata & 146.1  &  67.9 & \nodata & 0.053 & 0.026 & 0.006 & 9.999 & 9.999 & 9.999\\
ESO026-001  &  306.24542200 & -81.57592000 & 2 &  1.891 & 0.465 & 0.320 & 9.999 & 9.999 & 9.999\\
\nodata & 84.5  &  36.5 & \nodata & 0.046 & 0.110 & 0.013 & 9.999 & 9.999 & 9.999\\
ESO048-017  &  332.23184200 & -73.38194300 & 3 &  2.543 & 0.611 & 0.300 & 9.999 & 9.999 & 9.999\\
\nodata & 141.8 &   40.2 & \nodata & 0.064 & 0.092  & 0.009  & 9.999 & 9.999 & 9.999\\
ESO079-003  &    8.00970000 & -64.25241466 & 2 &  3.662 & 0.589 & 0.370 & 9.999 & 9.999 & 9.999\\
\nodata & 87.2  &  32.9 & \nodata & 0.100 & 0.071 & 0.009 & 9.999 & 9.999 & 9.999\\
ESO079-005  &   10.18214028 & -63.44262759 & 4 &  2.937 & 0.954 & 0.260 & 9.999 & 9.999 & 9.999\\
\nodata & 94.8  &  45.3 & \nodata & 0.055 & 0.060 & 0.006 & 9.999 & 9.999 & 9.999\\
ESO079-007  &   12.51660000 & -66.55376752 & 4 &  2.514 & 0.482 & 0.280 & 9.999 & 9.999 & 9.999\\
\nodata & 174.8 &   43.7 & \nodata & 0.067 & 0.019 & 0.005 & 9.999 & 9.999 & 9.999\\
ESO085-014  &   73.66965500 & -62.80207100 & 3 &  3.679 & 0.962 & 0.290 & 4.213 & 0.787 & 0.170\\
\nodata & 445.6 &  145.3 & \nodata & 0.079 & 0.041 & 0.004 & 0.106 & 0.074 & 0.005\\
ESO085-030  &   75.37484742 & -63.29272843 & 4 &  2.625 & 0.316 & 0.160 & 9.999 & 9.999 & 9.999\\
\nodata & 155.7 &   74.2 & \nodata & 0.106 & 0.016 & 0.003 & 9.999 & 9.999 & 9.999\\
ESO085-038  &   76.07969999 & -63.58132935 & 2 &  2.273 & 0.500 & 0.090 & 9.999 & 9.999 & 9.999\\
\nodata & 82.9  &  34.8 & \nodata & 0.059 & 0.100 & 0.004 & 9.999 & 9.999 & 9.999

\enddata
\tablecomments{CAS parameters and their uncertainties in the NUV and FUV for 
galaxies in our sample that were used for this study, with signtal-to-noise 
ratio S/N$_{NUV} \geq 75$. CAS values and their errors in the FUV are set 
to 9.999 for a null value (when S/N $< 75$).
Galaxy Type is listed as defined in Section 3. Coordinates are given in 
J2000. The entire dataset is available in the online edition.}
\end{deluxetable}

\noindent
\begin{deluxetable}{llllllllll}
\tabletypesize{\scriptsize}
\tablecolumns{10}
\tablewidth{0pt}
\tablecaption{Low S/N Measured CAS Parameters} 
\tablehead{
\colhead{Galaxy} & 
\colhead{R.A. (deg)} & 
\colhead{DEC (deg)} &
\colhead{Type} & 
\colhead{C$_{NUV}$} & 
\colhead{A$_{NUV}$} & 
\colhead{S$_{NUV}$} & 
\colhead{C$_{FUV}$} &
\colhead{A$_{FUV}$} &
\colhead{S$_{FUV}$}
\cr
\colhead{\nodata} &
\colhead{S/N$_{NUV}$} &
\colhead{S/N$_{FUV}$} &
\colhead{\nodata} &
\colhead{$\sigma_{C_{NUV}}$} &
\colhead{$\sigma_{A_{NUV}}$} &
\colhead{$\sigma_{S_{NUV}}$} &
\colhead{$\sigma_{C_{FUV}}$} &
\colhead{$\sigma_{A_{FUV}}$} &
\colhead{$\sigma_{S_{FUV}}$}
}
\startdata

2MASXJ11102621+5824080 & 167.60939026 &  58.40214920 & 2 &  3.066 & 0.189 & 0.200 &   2.853 & 0.247 & 0.130\\
\nodata & 88.3 &   30.4 & \nodata & 0.163 & 0.047 & 0.018  & 0.167 & 0.059  & 0.038\\
2MASXJ11554657+2956291 & 178.94429016 &  29.94148064 & 2 &  2.766 & 0.581 & 0.170 &   3.134 & -0.15 & 0.000\\
\nodata & 34.6 &   14.3 & \nodata & 0.095 & 0.105 & 0.010 & 0.151 & 0.210 & 0.000\\
2MASXJ16052284-0002177 & 241.34519958 &  -0.03841000 & 5 &  2.485 & -0.09 & -0.48 &   2.045 & 0.733 & -0.07\\
\nodata & 24.5 &    6.5 & \nodata & 0.109 & 0.190 & 0.020 & 0.127 & 0.270 & 0.009\\
ESO002-006             &   3.71070004 & -86.99308012 & 1 &  3.195 & 1.000 & 0.000 &   1.950 & 1.071 & 0.000\\
\nodata & 12.5 &   80.0 & \nodata & 0.173 & 0.174 & 0.000 & 0.144 & 0.302 & 0.000\\
ESO005-009             & 112.47719761 & -84.03832244 & 3 &  2.164 & 0.781 & 0.260 &   9.999 & 9.999 & 9.999\\
\nodata & 42.8 &    0.0 & \nodata & 0.092 & 0.088 & 0.009 & 9.999 & 9.999 & 9.999\\
ESO013-007             &  14.99880028 & -79.95686341 & 1 &  2.491 & -0.38 & 0.000 &   0.000 & 0.220 & 0.000\\
\nodata & 10.5 &    3.6 & \nodata & 0.243 & 0.197 & 0.000  & 0.369  & 0.381 & 0.000\\
ESO013-016             &  23.19899900 & -79.47364800 & 3 &  2.313 & 0.554 & 0.260 &   2.435 & 0.947 & 0.350\\
\nodata & 146.1 &    67.9 & \nodata & 0.053 & 0.026 & 0.006 & 0.051 & 0.048 & 0.013\\
ESO026-001             & 306.24542200 & -81.57592000 & 2 &  1.891 & 0.465 & 0.320 &   1.951 & 1.140 & 0.520\\
\nodata & 84.5 &   36.5 & \nodata & 0.046 & 0.110 & 0.013  & 0.043 & 0.134 & 0.022\\ 
ESO047-034             & 322.93289186 & -76.48059845 & 1 &  3.027 & 0.251 & 0.000 &   0.000 & -1.96 & 0.000\\
\nodata & 12.9 &    4.1 & \nodata & 0.249 & 0.257 & 0.000 & 0.369 & 0.206 & 0.000\\
ESO048-017             & 332.23184200 & -73.38194300 & 3 &  2.543 & 0.611 & 0.300 &   2.345 & 0.643 & 0.340\\
\nodata & 141.8 &   40.2 & \nodata & 0.064 & 0.092 & 0.009 & 0.061 & 0.105 & 0.016\\

\enddata
\tablecomments{CAS parameters and their uncertainties in the NUV and FUV for 
galaxies in our sample that have signtal-to-noise ratio S/N $< 75$ in one 
or both 
filters, which were rejected in our analysis. {\it We advise against 
using data with S/N $< 30$ in structure and morphology studies}. CAS values and 
their errors in the FUV are set to 9.999 for a null value (when no data 
exists).  Galaxy Type is listed as defined in Section 3. Coordinates are 
given in J2000. The entire dataset is available in the online edition.}
\end{deluxetable}

\noindent
\begin{deluxetable}{lllllll}
\tabletypesize{\scriptsize}
\tablecolumns{7}
\tablewidth{0pt}
\tablecaption{CAS UV Wavelength Dependence} 
\tablehead{
\colhead{Type Bin} & 
\colhead{$\delta_C$} & 
\colhead{$\sigma_C$} &
\colhead{$\delta_A$} & 
\colhead{$\sigma_A$} &
\colhead{$\delta_S$} & 
\colhead{$\sigma_S$}
}
\startdata
E/S0 & 0.45 & 0.11 & --0.15 & 0.03 & --0.03 & 0.02\\
Sa-Sc & 0.10 & 0.03 & --0.20 & 0.01 & --0.11 & 0.01\\
Sd-Im & 0.03 & 0.04 & --0.16 & 0.02 & --0.09 & 0.01\\
Pec & 0.01 & 0.06 & --0.14 & 0.03 & --0.07 & 0.02\\
\enddata
\tablecomments{The dependence of the CAS parameters on UV wavelength.
$\delta$ is defined as the median NUV CAS parameter 
for a given galaxy type bin minus the median FUV CAS parameter. $\sigma$
gives the combined uncertainty on the medians. 
}
\end{deluxetable}

\clearpage
\begin{figure}
\centerline{\includegraphics[width=15cm]{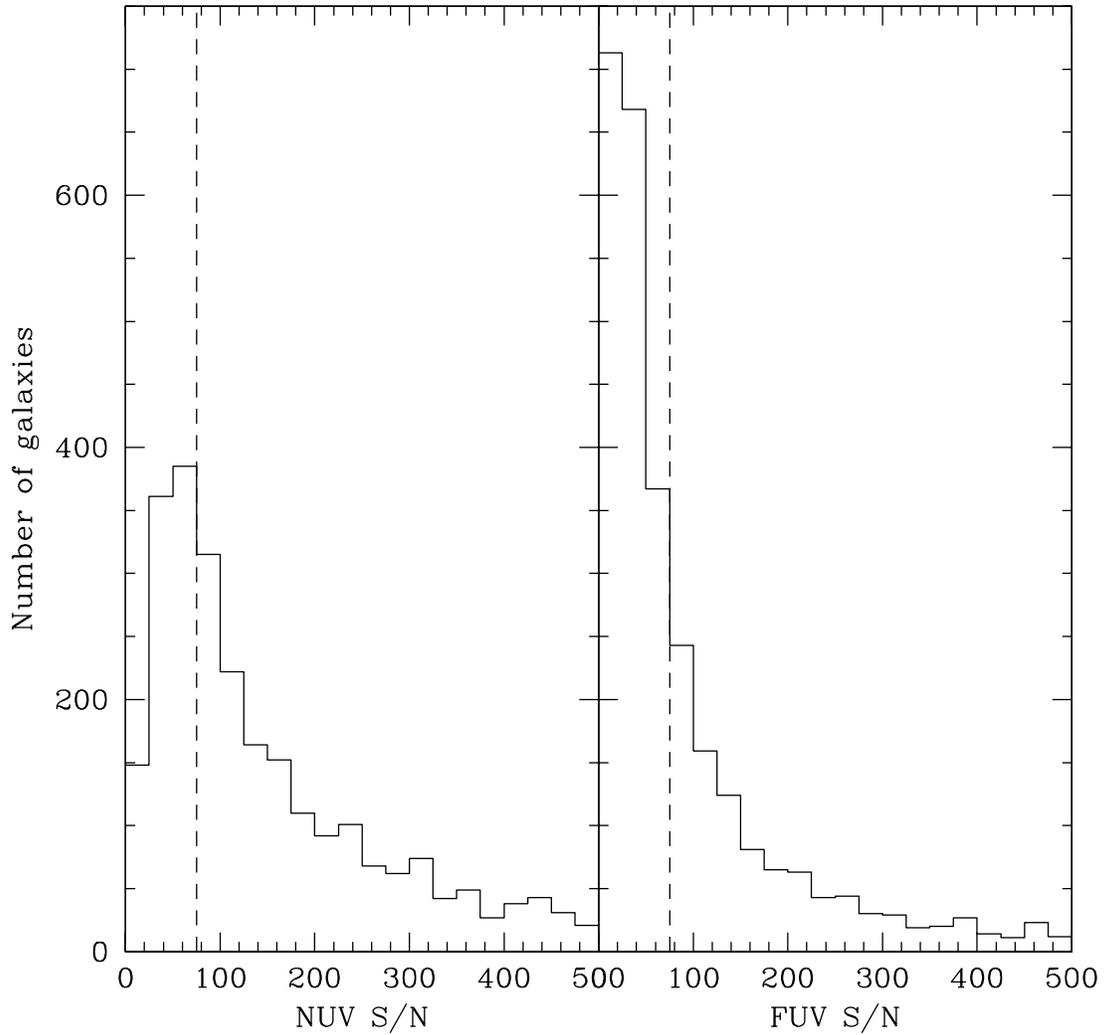}}
\figcaption[f1.eps]{\footnotesize{Histogram of signal-to-noise ratios (S/N) of GALEX
NUV (left panel) and FUV (right panel) galaxies in our initial sample.
Only galaxies with S/N $\geq 75$ were used in our analysis (dotted line).
Not shown here are 120 galaxies in FUV and 462 in NUV with S/N $> 500$.}
}
\end{figure}

\clearpage
\begin{figure}
\centerline{\includegraphics[width=15cm]{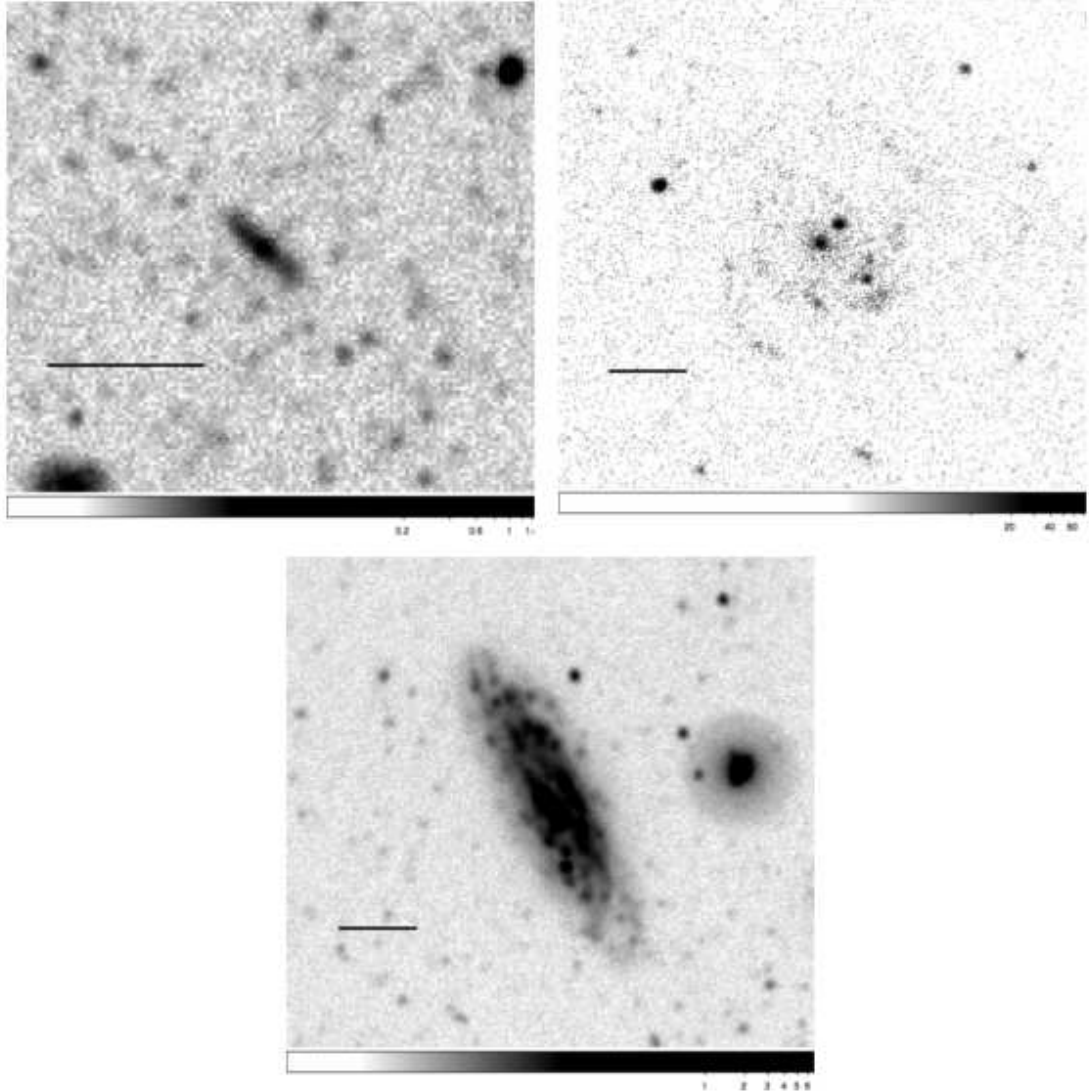}}
\figcaption[f2.eps]{\footnotesize{Examples of some galaxies of different quality
in signal-to-noise ratio in the GALEX NUV filter. The horizontal line
marks a length of $\sim 1$ arcminute (40 pixels).
{\bf Top left:} 2MASXJ11102621+5824080, with S/N of 88.
{\bf Top right:} ESO119-048, with S/N of 123. {\bf Bottom:}
ESO116-012, with S/N of 1330.}
}
\end{figure}

\clearpage
\begin{figure}
\centerline{\includegraphics[width=14cm]{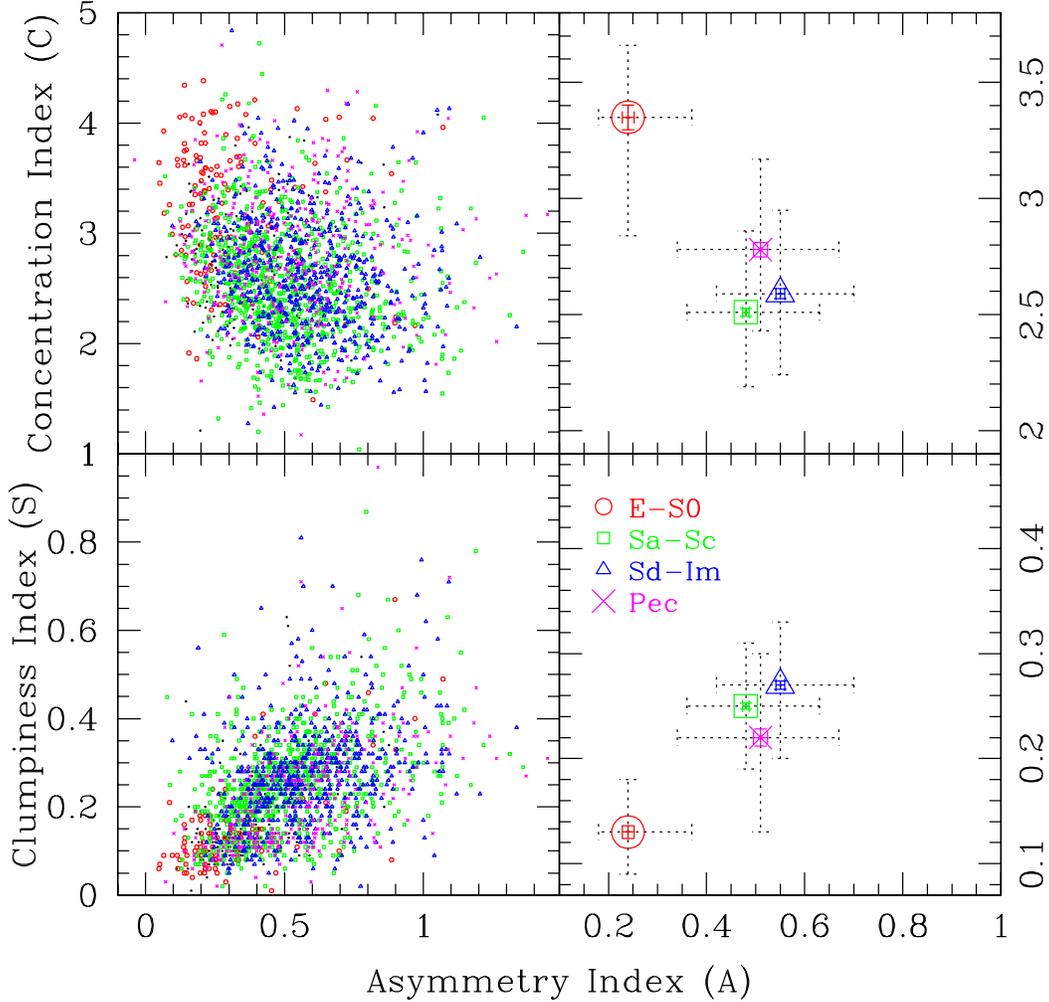}}
\figcaption[f3.eps]{\footnotesize{CAS parameter space for individual galaxies in the NUV
({\bf left}), and for the median values per galaxy type bin ({\bf right}). 
Dotted error bars indicate the 25--75\% quartile range. Solid error bars
indicate the error on the median, and are on the order of the size of the
data points, or smaller. Galaxies without an assigned
type are plotted with tiny solid black circles. A general trend exists of 
galaxies with increasingly higher asymmetry becoming less concentrated and 
more clumpy. Although there is a fair amount of overlap in CAS parameters
by galaxy type, the median values reveal that on average galaxies with
progressively earlier Hubble type have higher concentrations, lower
asymmetries, and lower clumpiness.}
}
\end{figure}

\clearpage
\begin{figure}
\centerline{\includegraphics[width=15cm]{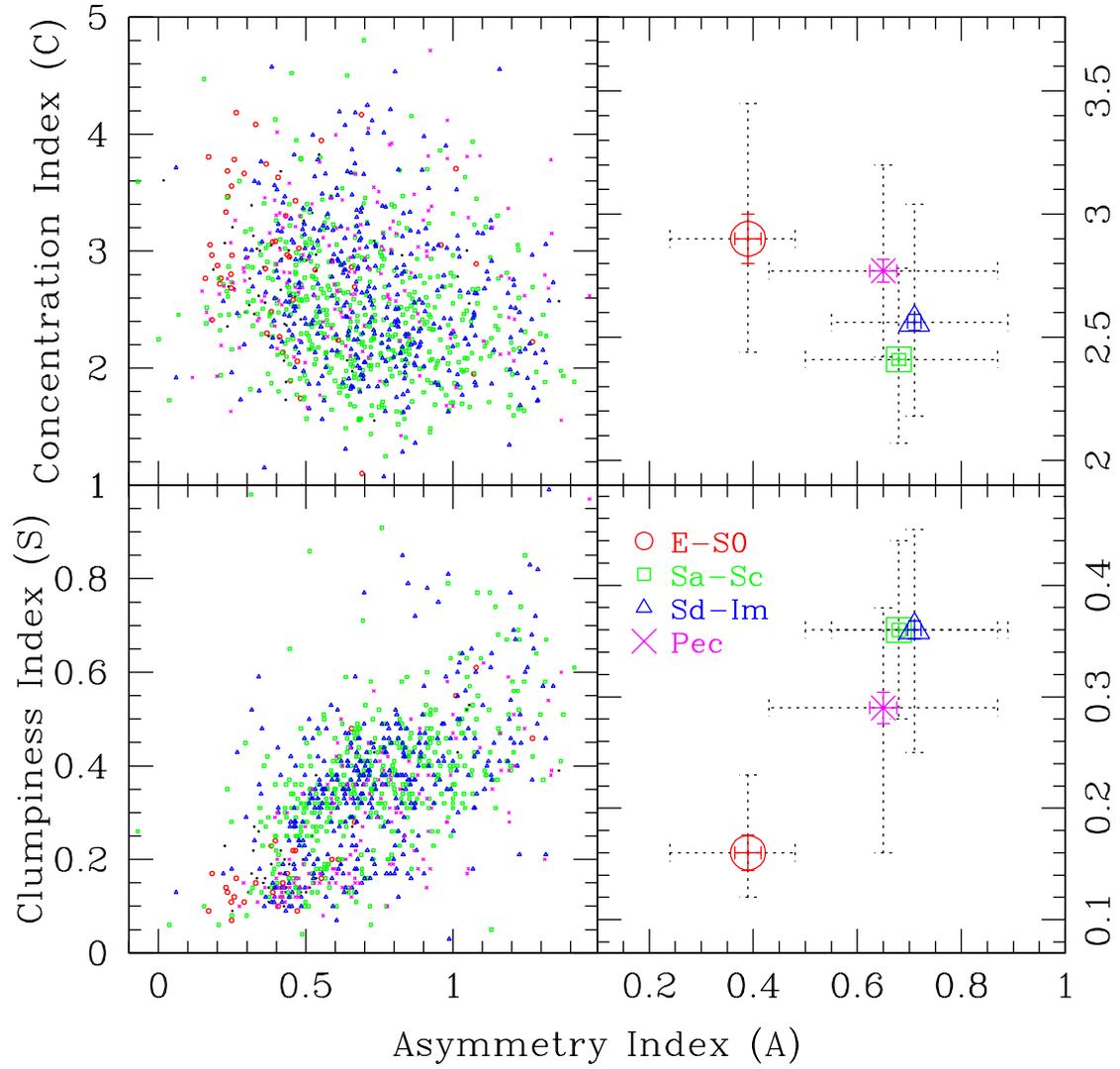}}
\figcaption[f4.eps]{\footnotesize{The same as in Figure 3, but with CAS parameters plotted
for galaxies in the FUV. Similar general trends are apparent in the FUV as 
in the NUV (Figure 3).}  
}
\end{figure}

\begin{figure}[!htb]
\centerline{\includegraphics[width=14cm]{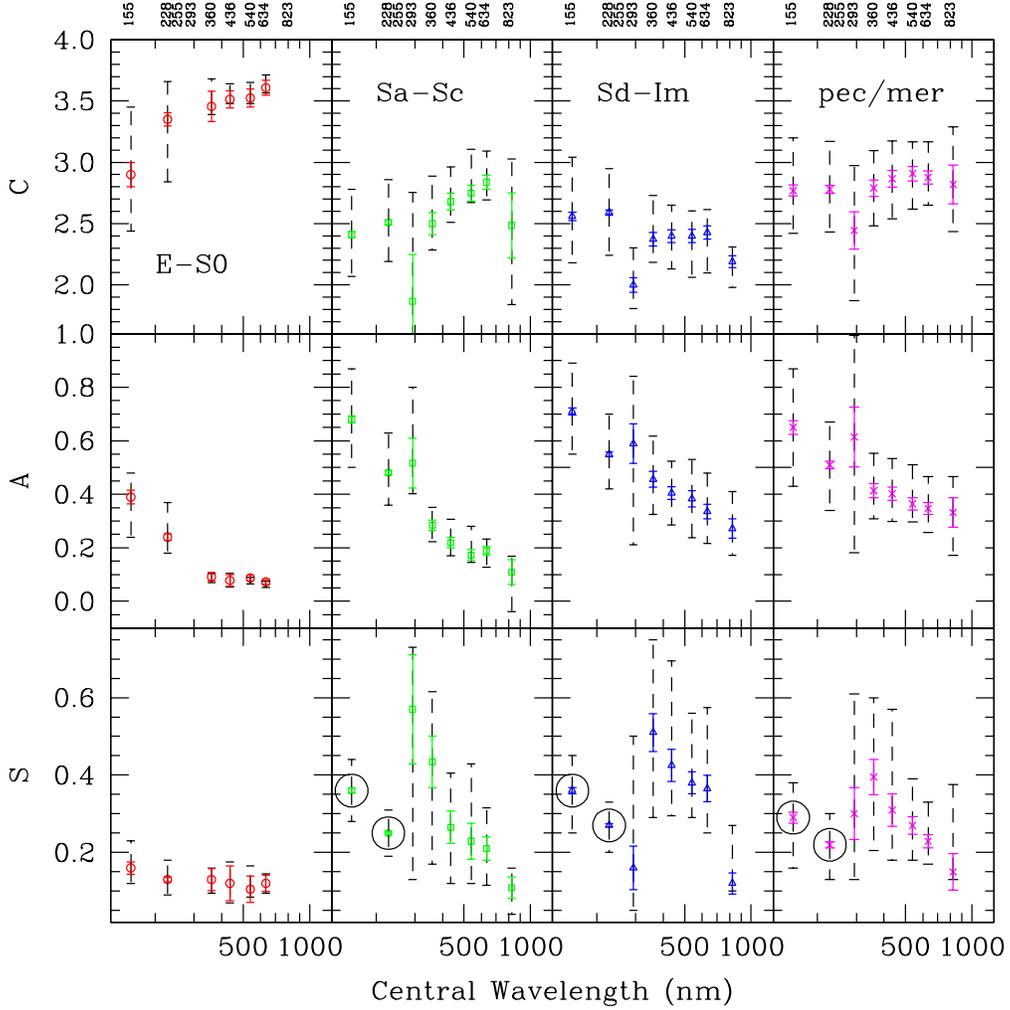}}
\figcaption[f5.eps]{\footnotesize{Median CAS values in each type bin as a function of
wavelength. Solid error bars indicate the error on the median, and dashed
error bars indicate the 25--75\% quartile range of the data. The GALEX data 
(two left-most data-points in each panel) is supplemented 
with HST and ground-based data. {\bf Circled data-points:} Low GALEX S values
are due to its poor (5 arcsec stellar FWHM) resolution. Disregarding this 
effect results 
in higher S values more in line with the trend displayed at longer 
wavelengths. In general:
galaxies are more asymmetric and more clumpy at
shorter rest-frame wavelengths, especially in the UV. This causes them 
to appear overall later-type than they would at longer wavelengths. 
Sa-Sc galaxies appear less concentrated at shorter
wavelengths, while Sd-Im and pec/mer galaxies do not exhibit
a strong concentration vs. wavelength relationship.
The effect in the UV is particularly
pronounced for elliptical/lenticular galaxies, which have little difference
in their CAS parameters in the near-UV through the near-IR, but appear
significantly less concentrated, more asymmetric, and slightly more clumpy
in the far-UV.}
}
\end{figure}

\end{document}